\documentclass[12pt]{article}
\usepackage{a4wide}
\usepackage{amssymb}
\usepackage{graphicx}
\begin{document}
{\renewcommand{\thefootnote}{\fnsymbol{footnote}}
\hfill  CGPG--03/3-4\\
\medskip
\hfill gr--qc/0303072\\
\medskip
\begin{center}
{\LARGE  Loop Quantum Cosmology, Boundary Proposals,\\ and Inflation}\\
\vspace{1.5em}
Martin Bojowald\footnote{e-mail address: {\tt bojowald@gravity.phys.psu.edu}}
and Kevin Vandersloot\footnote{e-mail address: {\tt kfvander@gravity.psu.edu}}
\\
\vspace{0.5em}
Center for Gravitational Physics and Geometry,\\
The Pennsylvania State
University,\\
104 Davey Lab, University Park, PA 16802, USA\\
\vspace{1.5em}
\end{center}
}

\setcounter{footnote}{0}

\newcommand{\case}[2]{{\textstyle \frac{#1}{#2}}}
\newcommand{\lP}{l_{\mathrm P}}

\newcommand{\md}{{\mathrm{d}}}
\newcommand{\tr}{\mathop{\mathrm{tr}}}
\newcommand{\sgn}{\mathop{\mathrm{sgn}}}

\newcommand*{\R}{{\mathbb R}}
\newcommand*{\N}{{\mathbb N}}
\newcommand*{\Z}{{\mathbb Z}}
\newcommand*{\Q}{{\mathbb Q}}
\newcommand*{\C}{{\mathbb C}}

\begin{abstract}
 Loop quantum cosmology of the closed isotropic model is studied with
 a special emphasis on a comparison with traditional results obtained
 in the Wheeler--DeWitt approach. This includes the relation of the
 dynamical initial conditions with boundary conditions such as the
 no-boundary or the tunneling proposal and a discussion of inflation
 from quantum cosmology.
\end{abstract}

\section{Introduction}

Traditionally, quantum cosmology has been studied in simple models
which have been obtained by a classical reduction of general
relativity to a system of finitely many degrees of freedom and a
subsequent application of quantum mechanical methods
\cite{DeWitt,Misner}. A corresponding full quantum theory of gravity,
let alone a relation of the models to it, has remained unknown beyond
a purely formal level. The most important question for quantum
cosmology, whether or not classical singularities are absent, has not
been answered positively in this approach. Instead, the classical
singularity has been removed by hand and substituted with some
proposals of intuitive initial conditions
\cite{DeWitt,nobound,tunneling}.

The situation has changed, however, with the advent of quantum
geometry (also called loop quantum gravity, see
\cite{Nonpert,Rov:Loops,ThomasRev}), a consistent canonical
quantization of full general relativity. It is possible to derive
cosmological models in a way analogous to the full theory
\cite{SymmRed} yielding loop quantum cosmology
\cite{cosmoI,IsoCosmo}. In this reduction, the main departure from the
traditional approach is the prediction that quantum Riemannian
geometry has a discrete structure.  The effects caused by the
discreteness are most important at small volume implying that the
structure close to the classical singularity is very different from
that of a Wheeler--DeWitt quantization which fails to resolve the
singularity problem in this regime. At large volume, however, the
discreteness leads only to small corrections and the traditional
approach is reproduced as an approximation.

Therefore, the main new results of loop quantum cosmology affect the
behavior at small volume: There is no singularity \cite{Sing}, initial
conditions for the wave function of a universe follow from the
evolution equation \cite{DynIn}, and the approach to the classical
singularity is modified by non-perturbative effects which imply an
inflationary period \cite{Inflation}. From the point of view of
initial conditions, the situation looks closest to DeWitt's original
proposal presenting, however, a well-defined generalization to other
models \cite{Scalar}.  So far all of these results have been derived and
studied in detail in the flat isotropic model only because a
non-vanishing intrinsic curvature leads to effects which complicate
the loop quantization. Recently, the methods of loop quantum cosmology
have been extended to homogeneous models with non-vanishing intrinsic
curvature \cite{Spin} such that now also a consistent quantization of
the closed isotropic model is available.

Since the traditional results have all been derived for the closed
model only, it is more important to compare with the effects of loop
quantum cosmology for that model. The general qualitative results
mentioned above remain true for the closed model, but some of them can
be different quantitatively. For instance, we will see that the
dynamical initial conditions resemble the no-boundary proposal more
closely than the tunneling proposal. This observation might appear
negative because the no-boundary proposal is widely perceived as being
unsupportive to large initial values of an inflaton field needed for a
sufficient amount of inflation. Here, however, we have an alternative
inflationary scenario based on the modified approach to classical
singularities which is also realized in the closed model. But due to
the intrinsic curvature the inflationary period does not necessarily
extend to arbitrarily small values of the scale factor which could
lead to insufficient inflation. As a new possibility we will therefore
look at the closed isotropic model as embedded in the anisotropic
Bianchi IX model. This leads to additional quantum modifications of
intrinsic curvature terms and of the classical behavior, implying a
second period of inflation at very small volume. With this more
general quantization the inflationary picture of the closed model is
more complicated but comparable to that of the flat
model.

\section{Loop Quantum Cosmology of the Closed Isotropic Model}

Quantum geometry and loop quantum cosmology are based on connection
and densitized triad variables \cite{AshVar,AshVarReell} which in the
isotropic case \cite{IsoCosmo,Bohr} are given by $c=\Gamma-\gamma K$
and $p$, respectively, where $K$ is the extrinsic curvature and the
parameter $\Gamma$ represents the intrinsic curvature with values
$\Gamma=0, \frac{1}{2}$ for the flat model and closed model
respectively. Both variables are canonically conjugate:
$\{c,p\}=\frac{1}{3}\gamma\kappa$ where $\kappa=8\pi G$ is the
gravitational constant and $\gamma$ is the Barbero--Immirzi parameter
\cite{AshVarReell,Immirzi} which is a positive real number and labels
classically equivalent formulations. The relation to the better known
ADM variables is given by $K=-\frac{1}{2}\dot{a}$ (extrinsic
curvature) and $|p|=a^2$ where $a$ is the scale factor. Since a triad
can have two different orientations, its isotropic component $p$ can
take both signs.

The dynamics is governed by the Hamiltonian constraint \cite{IsoCosmo}
\begin{eqnarray}
 H 
 &=& -6\gamma^{-2}\kappa^{-1}\left((c-2\Gamma)c+
   (1+\gamma^2)\Gamma^2\right) \sgn(p)\sqrt{|p|}\\
 &=& \left\{ \begin{array}{cl} -6\gamma^{-2}\kappa^{-1}c^2
     \sgn(p)\sqrt{|p|} & \mbox{for the flat model} \\
     -6\gamma^{-2}\kappa^{-1} (c^2-c+\case{1}{4}(1+\gamma^2))
     \sgn(p)\sqrt{|p|} & \mbox{for the closed model}
   \end{array}\right.
\end{eqnarray}
or in terms of the extrinsic curvature
\begin{eqnarray}
 H &=& -6\kappa^{-1}(K^2+\Gamma^2) \sgn(p)\sqrt{|p|}\nonumber\\
 &=& \left\{\begin{array}{cl} -\case{3}{2}\kappa^{-1}\dot{a}^2a &
     \mbox{flat} \\ -\case{3}{2}\kappa^{-1}(\dot{a}^2+1)a &
     \mbox{closed} \end{array}\right.
\end{eqnarray}
which in the form of the constraint equation $H+H_{\rm matter}(a)=0$
with the matter Hamiltonian $H_{\rm matter}(a)$ yields the usual
Friedmann equation
\begin{equation} \label{Friedmann}
 a^{-3}H+\rho_{\rm matter}(a)=0\,.
\end{equation}

In quantum geometry holonomies of the connection together with flux
variables associated with the densitized triad are promoted to basic
operators. One usually works in the connection representation where
states are functionals on the infinite-dimensional space of
connections via holonomies. After reducing to isotropic variables only
one gauge invariant connection component $c$ remains and states in the
connection representation are functions of this single parameter. An
orthonormal basis is given by the states
\begin{equation}
  \langle c|n\rangle:=\frac{\exp(\case{1}{2}inc)}{\sqrt{2}\sin\case{1}{2}c}
\end{equation}
labeled by an integer $n$. The states $|n\rangle$ are eigenstates of
the basic derivative operator $\hat{p}$ which quantizes the isotropic
triad component
\begin{equation}
 \hat{p}|n\rangle=\case{1}{6}\gamma\lP^2n|n\rangle
\end{equation}
where the Planck length $\lP=\sqrt{\kappa\hbar}$ appears. We will
later mainly use the volume operator which also has the states
$|n\rangle$ as eigenstates:
\begin{equation} \label{vol}
 \hat{V}|n\rangle= V_{\frac{1}{2}(|n|-1)}|n\rangle=
 (\case{1}{6}\gamma\lP^2)^{\frac{3}{2}} \sqrt{(|n|-1)|n|(|n|+1)}
   \,|n\rangle\,.
\end{equation}
Composite operators will be constructed from the volume operator
together with multiplication operators
\begin{eqnarray}
 \cos(\case{1}{2}c) |n\rangle &=& \case{1}{2}\left(\exp(\case{1}{2}ic)+
   \exp(-\case{1}{2}ic)\right) \frac{\exp(\case{1}{2}inc)}{\sqrt{2}
   \sin(\case{1}{2}c)}= \case{1}{2}(|n+1\rangle+|n-1\rangle) \label{cos}\\
 \sin(\case{1}{2}c) |n\rangle &=& -\case{1}{2}i\left(\exp(\case{1}{2}ic)-
   \exp(-\case{1}{2}ic)\right) \frac{\exp(\case{1}{2}inc)}{\sqrt{2}
   \sin(\case{1}{2}c)}= -\case{1}{2}i(|n+1\rangle-
 |n-1\rangle)\,. \label{sin}
\end{eqnarray}

An example is presented by the inverse scale factor operator which is
needed, e.g., in order to quantize the kinetic part of matter
Hamiltonians. Since the triad and volume operators have a discrete
spectrum containing the value zero, they do not have a densely defined
inverse which would be necessary for a well-defined operator. However,
there are methods in quantum geometry \cite{QSDV} which allow the
classically divergent inverse scale factor to be turned into a
well-defined operator \cite{InvScale}. To that end, one has to rewrite
it classically as, e.g.,
\[
 a^{-1}= 6\gamma^{-1}\kappa^{-1}\{c,V^{\frac{1}{3}}\}=
 12i\gamma^{-1}\kappa^{-1}
 e^{\frac{1}{2}ic}\{e^{-\frac{1}{2}ic},V^{\frac{1}{3}}\}
\]
using the symplectic structure. On the right hand side an inverse of
the volume does not appear anymore, and it can easily be quantized by
using the multiplication operators (\ref{cos}), (\ref{sin}), the
volume operator (\ref{vol}), and turning the Poisson bracket into a
commutator. As a result, the classical divergence of $a^{-1}$, which
is also present in Wheeler--DeWitt quantizations where $a$ simply
becomes a multiplication operator, is cut off at small volume and the
inverse scale factor operator is finite. We also note that rewriting a
classical quantity in the way above introduces new possibilities
for quantization ambiguities. For instance, we wrote the connection
component $c$ as an exponential which can appear with an arbitrary
integer power. This integer determines at which volume the classical
divergence is cut off \cite{Ambig}, an effect which will be used and
discussed in more detail later.

Another example for a composite operator which can be built from the
volume and multiplication operators is the Hamiltonian constraint. We
first recall the result in the flat isotropic case where methods used
in the full theory \cite{QSDI} lead to the constraint operator
\cite{cosmoIII,IsoCosmo}
\begin{equation}\label{Hamflat}
 \hat{H}_{\rm flat}= 96i(\gamma^3\kappa\lP^2)^{-1}
 \sin^2(\case{1}{2}c)\cos^2(\case{1}{2}c) \left(\sin(\case{1}{2}c)\hat{V}
 \cos(\case{1}{2}c)- \cos(\case{1}{2}c) \hat{V} \sin(\case{1}{2}c)\right)
\end{equation}
with action
\begin{equation}
 \hat{H}_{\rm flat}|n\rangle= 3(\gamma^3\kappa\lP^2)^{-1}\sgn(n)
 \left(V_{\frac{1}{2}|n|}- V_{\frac{1}{2}|n|-1}\right) (|n+4\rangle
 -2|n\rangle+ |n-4\rangle)\,.
\end{equation}
As usual, the constraint equation $\hat{H}|s\rangle=-\hat{H}_{\rm
matter}|s\rangle$ turns into an evolution equation after transforming
to a triad representation. In a triad representation the state
$|s\rangle$ is represented by the coefficients $s_n(\phi)$ which
appear in a decomposition of the state $|s\rangle=\sum_n
s_n(\phi)|n\rangle$ in terms of triad eigenstates $|n\rangle$. Here,
we also write the dependence of the wave function on a matter value
$\phi$ which we do not need to specify further for our purposes. The
matter Hamiltonian $\hat{H}_{\rm matter}$ acts on the
$\phi$-dependence of the state, but via metric components also on the
gravitational part with label $n$.  Since the triad has discrete
spectrum in loop quantum cosmology, the evolution equation is a
difference equation in the discrete internal time $n$, rather than a
second order differential equation as in a Wheeler--DeWitt
quantization. Still, for large volume $n\gg1$ and small extrinsic
curvature the difference equation can be well approximated by a
differential equation such that the Wheeler--DeWitt approach and thus
the semiclassical limit is reproduced at large volume
\cite{SemiClass}. At small volume close to the classical singularity,
however, there are large corrections which must not be ignored.

In the flat model as discussed above, the intrinsic curvature is
always zero such that small extrinsic curvature at large volume
implies small total curvature and we expect almost classical
behavior. In the closed model, on the other hand, the intrinsic
curvature represented by the parameter $\Gamma=\frac{1}{2}$ is
constant and never small. A general consistent loop quantization of
homogeneous models \cite{HomCosmo} can be obtained by subtracting
$\Gamma$ from the connection component $c$ such that we obtain the
constraint operator
\begin{equation}
 \hat{H}=96i(\gamma^3\kappa\lP^2)^{-1}
 \left(\sin^2\left(\case{1}{2}(c-\case{1}{2})\right)
   \cos^2\left(\case{1}{2}(c-\case{1}{2})\right)+
   \case{1}{16}\gamma^2\right) \left(\sin(\case{1}{2}c)\hat{V}
 \cos(\case{1}{2}c)- \cos(\case{1}{2}c) \hat{V} \sin(\case{1}{2}c)\right)
\end{equation}
with action
\begin{equation}
 \hat{H}|n\rangle= 3(\gamma^3\kappa\lP^2)^{-1}\sgn(n)
 \left(V_{\frac{1}{2}|n|}- V_{\frac{1}{2}|n|-1}\right) (e^{-i}|n+4\rangle
 -(2+\gamma^2)|n\rangle+ e^i|n-4\rangle)\,.
\end{equation}
Transforming to the triad representation results in the difference
equation
\begin{eqnarray} \label{Diff}
 && \sgn(n+4)(V_{|n+4|/2}-V_{|n+4|/2-1})e^is_{n+4}(\phi)-
 (2+\gamma^2)\sgn(n)(V_{|n|/2}-V_{|n|/2-1})s_n(\phi)\nonumber\\
 &&+
 \sgn(n-4)(V_{|n-4|/2}-V_{|n-4|/2-1})e^{-i}s_{n-4}(\phi)\nonumber\\
 &=& -\case{1}{3}\gamma^3\kappa\lP^2\hat{H}_{\rm matter}(n)s_n(\phi)
\end{eqnarray}
where the reduced matter Hamiltonian $\hat{H}_{\rm matter}(n)$ defined
by $\hat{H}_{\rm matter}|n\rangle \otimes |\phi\rangle=: |n\rangle
\otimes \hat{H}_{\rm matter}(n) |\phi\rangle$ acts only on the
$\phi$-dependence of the wave function. One can easily check that it
is possible to evolve through the classical singularity at $n=0$ in
the same way as in the flat case \cite{Sing}; thus, the singularity is
absent in the loop quantization.

In order to derive a continuum approximation at large volume we have
to find a continuous wave function $\psi(p,\phi)$ related to the
discrete wave function $s_n(\phi)$ which does not vary strongly at
small scales and solves an approximating differential equation derived
from (\ref{Diff}). Since a good continuum limit is a prerequisite for
a correct classical limit, a wave function allowing such an
interpolation is called pre-classical \cite{DynIn}. In the flat case
the wave function $s_n$ itself turned out to lead to the pre-classical
solutions, while general considerations show that this cannot be the
case in the presence of large intrinsic curvature \cite{Spin}. In this
case the wave function $\tilde{s}=\exp(3i\Gamma\hat{p}/\gamma\lP^2)s$
with coefficients
\begin{equation}
 \tilde{s}_n(\phi)=e^{\frac{1}{2}in\Gamma}s_n(\phi)=
 e^{\frac{1}{4}in}s_n(\phi)
\end{equation}
where the phase factor cancels small-scale oscillations in $s_n$ has
to be used for a continuum limit. As in \cite{SemiClass} for the flat
case it can then be verified that the approximating differential
equation for $n\gg1$ is
\begin{equation} \label{WdW}
\case{1}{2} \left(\case{4}{9}\lP^4\frac{\partial^2}{\partial p^2}-
  1\right) \sqrt{|p|}\psi(p,\phi)= -\case{1}{3}\kappa \hat{H}_{\rm
  matter}(p)\psi(p,\phi)
\end{equation}
where $\psi(p,\phi)$ is an interpolation of the pre-classical
$\tilde{s}_n(\phi)$. The left hand side can be written as
$-\frac{1}{2}(4\hat{K}^2+1) a\psi(a,\phi)$ which shows that the
Wheeler--DeWitt equation as a Schr\"odinger-like quantization of
(\ref{Friedmann}) is reproduced at large volume. The ordering is fixed
in (\ref{WdW}) and follows from the non-singular difference equation
of loop quantum cosmology. For de Sitter space with cosmological
constant $\Lambda$ we obtain the general solution
\[
 \sqrt{|p|}\psi(p)=A{\mathrm{Ai}}\left(-(\case{3}{2}\Lambda)^{\frac{1}{3}}
 (p-\case{3}{2}\Lambda^{-1})\right)+
 B{\mathrm{Bi}}\left(-(\case{3}{2}\Lambda)^{\frac{1}{3}}
 (p-\case{3}{2}\Lambda^{-1})\right)
\]
in terms of Airy functions. The wave function $\psi(p)$ is either zero
in $p=0$, if and only if
$A{\mathrm{Ai}}\left((\case{3}{2})^{\frac{4}{3}}\Lambda^{-\frac{2}{3}}\right)=
-B{\mathrm{Bi}}\left((\case{3}{2})^{\frac{4}{3}}\Lambda^{-\frac{2}{3}}\right)$,
or diverges there.

\section{Dynamical Initial Conditions}

When we approach the classical singularity at $n=0$ by going to
smaller values of $n$, we have to take into account discrepancies
between the Wheeler--DeWitt equation and the difference equation of
loop quantum cosmology. As a difference equation, the exact constraint
equation (\ref{Diff}) can be used as a recurrence relation which
determines the wave function starting from initial values at some
finite positive values of $n$. We can find values $s_n$ for smaller
$n$ as long as the coefficient $V_{|n-4|/2}-V_{|n-4|/2-1}$ of lowest
order in (\ref{Diff}) does not vanish. However, it does vanish if and
only if $n=4$ such that we cannot fix $s_0(\phi)$ in this way. As in
\cite{Sing} this does not mean that there is a singularity because we
can evolve through $n=0$ and find all values of the wave function for
negative internal time $n$. But the part of the constraint equation
for $n=4$ has to be satisfied. Instead of fixing $s_0$ in terms of the
initial data, it implies a linear relation between $s_4$ and $s_8$
which have already been determined in previous steps of the
recurrence. Therefore, the constraint equation imposes implicitly a
linear consistency condition on the initial data which serves as an
initial condition. Since it is not imposed separately but follows from
the evolution equation, it is called a dynamical initial condition
\cite{DynIn}. Note that it only gives us partial information and does
not fix the wave function completely in the presence of matter fields.

While the value $s_0$ of the wave function at the classical
singularity is undetermined and it thus is impossible to formulate the
dynamical initial conditions as conditions at $n=0$, they imply
effectively that a pre-classical wave function has to approach zero
close to $n=0$. In this sense, it is similar to DeWitt's initial
condition which requires the wave function to vanish at the classical
singularity. In contrast to that condition, however, the dynamical
initial condition is well-posed in the discrete context (see, e.g.,
\cite{Scalar}).

For the closed model we can now also compare the dynamical initial
condition with other traditional proposals, most importantly the
no-boundary \cite{nobound} and the tunneling proposal
\cite{tunneling}, which have been discussed only in this case. To do
that we first consider the approximate differential equation for the
Hamiltonian constraint (\ref{WdW}) and rewrite it in terms of the
scale factor $a$ and in the ``de Sitter'' approximation where the
scalar field potential is approximated by a (cosmological) constant
$V$ and the scalar field dependence is ignored
\begin{equation} \label{deSitter}
\left[a \frac{\md}{\md a} \frac{1}{a} \frac{\md}{\md a} - 
      \frac{a^2}{\lP^4} \left( 9 - 6\kappa V a^2\right) \right] 
      a \Psi(a) = 0 \,.
\end{equation}
This represents an ordering of the Hamiltonian constraint that has not
previously been considered (compare, e.g., the general analysis in
\cite{Konto}).  If we write $\tilde{\Psi}=a\Psi$ we see that in order
for $\Psi$ to be bounded for small $a$ we demand that $\tilde{\Psi}$
approach zero. For $\kappa a^2V\ll 1$ the solution to (\ref{deSitter})
satisfying $\tilde{\Psi}(0)=0$ is $AaI_{1/2} (\frac{3a^2}{2\lP^2})$
where $I_{1/2}$ is the modified Bessel function and $A$ is an
arbitrary constant. When matched to the WKB solutions in the
exponential region, this solution picks out the exponentially
increasing WKB mode thus resembling the no-boundary wave function. We
note however, that in the limit of small $a$ $\Psi$ approaches zero
satisfying DeWitt's initial condition.

\begin{figure}[ht]
\begin{center}
 \includegraphics[width=12cm,height=8cm,keepaspectratio]{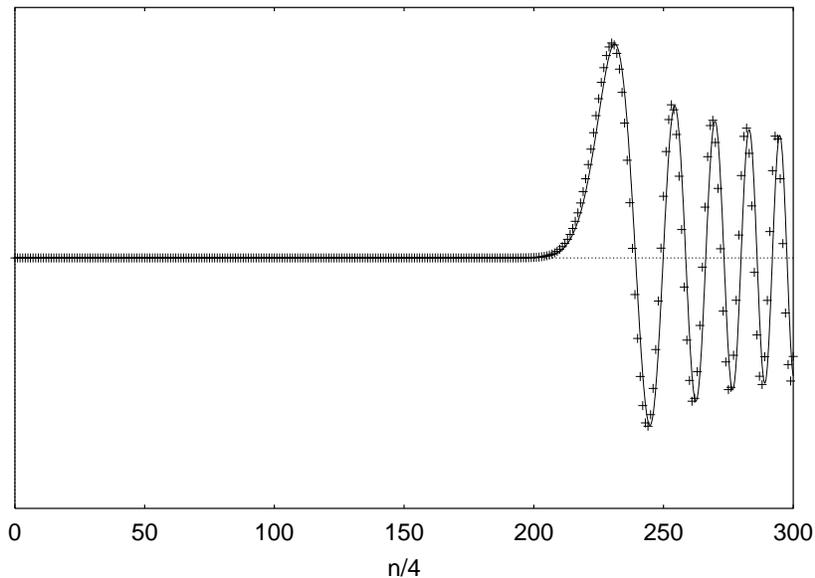}
\end{center}
\caption{Pre-classical solution $(V_{|n|/2}-V_{|n|/2-1})\tilde{s}_n$
  of the discrete equation (\ref{Diff}) compared to a solution
  $\sqrt{p}\psi(p)$ of the Wheeler--DeWitt equation (\ref{WdW}) such
  that $\psi$ is regular at $p=0$ (solid line) corresponding to de
  Sitter space with $\kappa\Lambda=10^{-2}\lP^{-2}$.}
\label{WaveFunc}
\end{figure}

The previous analysis in terms of $\Psi(a)$ shares the same flaws of
the traditional approach to quantum cosmology in that conditions on
the wave function are specified near the classical singularity where
we might expect to see large corrections due to the effects of
discrete space. The evolution equation (\ref{Diff}) needs to be solved
in order to determine if the discrete effects modify the
analysis. Since the dynamical initial condition forces the discrete
wave function to approach zero for small $a$, the wave function
increases for growing $a$ before it starts to oscillate. Similarly to
the previous analysis, the wave function exponentially increases in
the classically forbidden region.  A numerical solution to
(\ref{Diff}) with a cosmological constant exhibits this behavior
(Fig.~\ref{WaveFunc}) reinforcing the previous analysis and indicating
that the dynamical initial condition imposes conditions similar to the
no-boundary proposal.

\section{Inflation}

It has been speculated that the no-boundary proposal does not predict
large enough initial values for an inflaton field which suggests that
a sufficient amount of inflation cannot be realized
\cite{tunnelinginfl}. If correct, the same conclusion would apply for
a wave function satisfying the dynamical initial condition. However,
loop quantum cosmology presents an alternative mechanism of inflation
\cite{Inflation} which does not necessarily require an inflaton
field. This scenario exploits quantum modifications of the classical
equations of motion implied by the discrete formulation. There are
several types of corrections, but we can focus on the non-perturbative
effect which results from a quantization of inverse metric components
since it implies the most drastic changes. The effect contains an
ambiguity parameter and by choosing it to be large the
non-perturbative modification extends into the semiclassical regime.
The wave function can then be approximated as a wave packet following
the effective classical equations of motion which we will discuss now.

Eigenvalues of a density operator $\hat{d}_j=\widehat{a^{-3}}_j$, which
quantizes the classical density $a^{-3}$ in the kinetic part of a
matter Hamiltonian, can be computed explicitly \cite{Ambig} for
arbitrary values of the ambiguity parameter $j$ which is a
half-integer. Here we only need an approximate expression for the
eigenvalues
\[
 d_{j,n}\sim (\case{1}{6}\gamma\lP^2n)^{-\frac{3}{2}} p(n/2j)^6
\]
where the function
\begin{equation} \label{p}
 p(q)=\case{8}{77}
 q^{\frac{1}{4}} \left(
 7\left((q+1)^{\frac{11}{4}}- |q-1|^{\frac{11}{4}}\right)- 11q\left(
 (q+1)^{\frac{7}{4}}- \sgn(q-1)|q-1|^{\frac{7}{4}}\right) \right)
\end{equation}
approaches one at large values but incorporates non-perturbative
corrections for small $n<2j$. In effective classical equations of
motion the modified density will appear as a continuous function
\begin{equation} \label{dj}
 d_j(a)=a^{-3}p(3a^2/\gamma\lP^2j)^6\sim \left\{\begin{array}{cl}
     \case{12^6}{7^6} (\case{1}{3}\gamma\lP^2j)^{-\frac{15}{2}} a^{12} &
     \mbox{for }a^2\ll\case{1}{3}\gamma\lP^2j \\ a^{-3} &\mbox{for
       }a^2\gg\case{1}{3}\gamma\lP^2j \end{array}\right.
\end{equation}
derived using $a^2\sim\case{1}{6}\gamma\lP^2n$. The precise form of
the behavior for small $a$ is subject to quantization ambiguities, but
qualitatively it always has the above form. Additionally, the power of
$a$ in the small-$a$ approximation is usually high as in the example
used here.

In particular, $d_j(0)=0$ and $d_j(a)$ increases as a function of $a$
at small volume in contrast to the classical divergence. Consequently,
the effective matter Hamiltonian, e.g., $H_{\rm matter}^{\rm eff}(a)=
\frac{1}{2} d_{j_{\phi}}(a)p_{\phi}^2+a^3V(\phi)$ for a scalar, always
satisfies
\[
 \lim_{a\to 0}a^{-1}H_{\rm matter}^{\rm eff}(a)=0
\]
even taking into account the kinetic term which would diverge
classically. We can view the Friedmann equation $\dot{a}^2+V(a)=0$ as
describing a classical motion in the effective potential
\begin{equation}
 V(a)=1-\case{2}{3}\kappa a^{-1} H_{\rm matter}^{\rm eff}(a)
\end{equation}
where we ignore the dependence on the matter field.  Because the
modified density is increasing at small $a$ just like a potential
term, the derivative $V'(a)$ of the effective potential is negative
for small $a$. The effective classical equation of motion then implies
\[
 \ddot{a}=-\case{1}{2}V'(a)>0 \qquad\mbox{for small $a$}
\]
such that the effective classical evolution at small $a$ is
inflationary. An inflaton field and therefore large initial values are
not required for this kind of inflation which is purely due to quantum
geometry effects.

It is, however, necessary that there be a large enough overlap between
the region where the effective potential decreases and the classically
allowed region where the effective potential is negative. In this
intersection we can use the effective classical equations of motion
and obtain an amount of inflation $a_{\rm f}/a_{\rm i}$ given by the
initial scale factor $a_{\rm i}$ where the classically allowed region
starts and the final scale factor $a_{\rm f}$ where the modified
increasing density goes over into the standard density $a^{-3}$. For
the flat model, the entire region of positive $a$ was classically
allowed, and $a_{\rm i}$ can be arbitrarily small, only restricted by
the eventual breakdown of the effective classical approximation. In
the closed model, however, the non-zero intrinsic curvature leads to a
hill in the effective potential at small $a$ such that the classically
allowed region starts at a positive value of the scale factor. An
approximate expression for the ratio $a_{\rm f}/a_{\rm i}$ can be
derived in the case of a free massless scalar field.  Using the
small-$a$ approximation of $d_j$ in (\ref{dj}), the classical turning
point corresponds to where the effective potential equals zero. This
gives
\begin{equation}
a^{11}_{\rm i} \approx \frac{3}{\kappa} \frac{7^6}{12^6} \left(\frac {\gamma
\lP^2 j_{\phi}}{3} \right)^{15/2} \frac{1}{p_{\phi}^2}
\end{equation}
where $p_{\phi}$ is the conjugate scalar field momentum. The
inflationary regime ends where $d_{j_{\phi}}(a)$ takes on its maximum
value which corresponds to $a_{\rm f}\approx\sqrt{\gamma \lP^2
j_{\phi}/3}$ \cite{Ambig}. For practical values of the parameters the
inflationary region is small and getting large amounts of inflation
would require large values of $p_{\phi}$.

\subsection{Suppression of Intrinsic Curvature}

With the quantization of the closed model presented so far it is hard
to get a large amount of inflation because the intrinsic curvature
introduces a potential hill which prohibits the scale factor from
attaining very small values classically. The divergence of the kinetic
part of matter Hamiltonians, which could cancel the potential hill, is
cut off by quantum geometry effects. Since the inverse scale factor in
isotropic cosmologies is related to the extrinsic curvature, this can
also be interpreted as an extrinsic curvature cut-off. Because
intrinsic and extrinsic curvature belong to the same geometrical
object in a covariant treatment, one could expect that there is a
similar effect which suppresses the intrinsic curvature represented by
the spin connection $\Gamma=\frac{1}{2}$. This can in fact be realized
by viewing the closed isotropic model as being embedded in the
anisotropic Bianchi IX model. An anisotropic model has three
independent triad components $p^I$ which determine the intrinsic
curvature via the spin connection
$\Gamma_1=\frac{1}{2}(p^2/p^3+p^3/p^2-p^2p^3/(p^1)^2)$ and analogous
components $\Gamma_2$ and $\Gamma_3$ (for Bianchi IX,
\cite{HomCosmo}). The isotropic model can be obtained by fixing
$p^1=p^2=p^3=p$ such that $\Gamma_I=\frac{1}{2}=\Gamma$. The
Hamiltonian constraint of the Bianchi IX model \cite{Spin} requires a
quantization of the spin connection which has to be a well-defined
operator. Since the components $\Gamma_I$ contain inverse powers of
the triad, we have to use inverse triad operators introducing another
ambiguity parameter $j_{\Gamma}$ which can be different from the
parameter $j_{\phi}$ used when quantizing the density $d=a^{-3}$
appearing in a scalar matter Hamiltonian. After reducing the quantized
Bianchi IX spin connection to isotropy we obtain, with $p(q)$ of
(\ref{p}), eigenvalues
\[
 \Gamma_{j_{\Gamma},n}=\case{1}{2}\left(2p(n/2j_{\Gamma})^2-
			                p(n/2j_{\Gamma})^4\right)
\]
replacing $\Gamma=\frac{1}{2}$ in the constraint equation. When $n$ is
small, the effective inverse triad components decrease such that
$\Gamma_{j_{\Gamma}}$ becomes smaller than $\frac{1}{2}$ for small
volume and approaches zero at the classical singularity. For $n\gg
2j_{\Gamma}$ we have $\Gamma_{j_{\Gamma},n}\sim\frac{1}{2}$. In fact,
we do have a suppression of intrinsic curvature just as the extrinsic
curvature is suppressed.

\begin{figure}[ht]
\begin{center}
 \includegraphics[width=12cm,height=8cm,keepaspectratio]{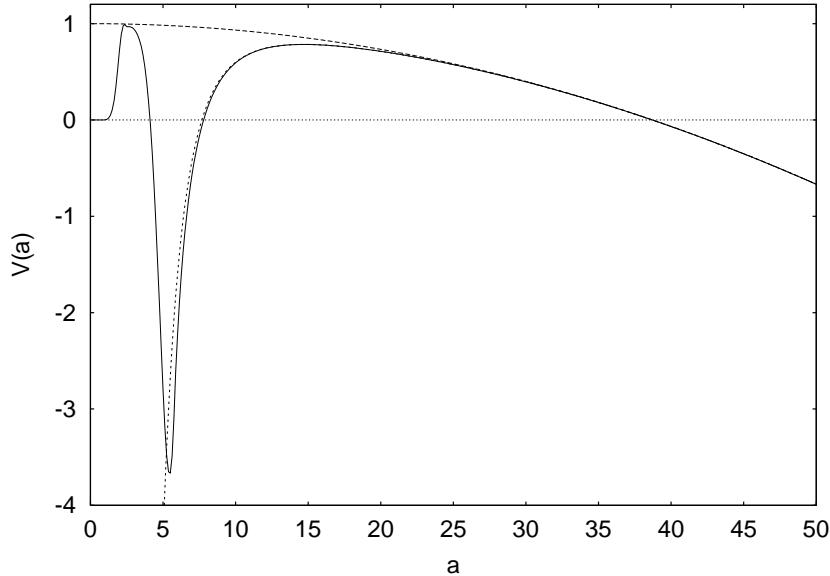}
\end{center}
\caption{Example of an effective potential $V_{\Gamma}(a)$ ($a$ in
  multiples of the Planck length) for a massless scalar with zero
  potential and a cosmological constant
  $\kappa\Lambda=10^{-3}\lP^{-2}$. The ambiguity parameters are
  $j_{\Gamma}=20$, $j_{\phi}=100$, and the scalar momentum is
  $\sqrt{\kappa}p_{\phi}=100\lP^2$. For larger values of $a$ the
  potential continues to decrease since the cosmological constant term
  dominates. The dashed lines are the effective potential without
  $\Gamma$-suppression and kinetic term (approaching one for $a=0$),
  and without $\Gamma$-suppression and with a standard kinetic term
  (which diverges for $a=0$).}
\label{EffPot}
\end{figure}

As a consequence, the potential hill in the effective potential
shrinks because the intrinsic curvature term which equaled one before
is not constant anymore and approaches zero: $\lim_{a\to
0}V_{\Gamma}(a)=0$ rather than one (see Fig.~\ref{EffPot}), where
\[
 V_{\Gamma}(a)=4\Gamma_{\rm eff}^2(a)-\case{2}{3}\kappa a^{-1}H_{\rm
 matter}^{\rm eff}(a)
\]
and
\[
 \Gamma_{\rm eff}(a)=\case{1}{2}\left(2p(3a^2/\gamma\lP^2
 j_{\Gamma})^2- p(3a^2/\gamma\lP^2 j_{\Gamma})^4\right)\,.
\]

\begin{figure}[ht]
\begin{center}
 \includegraphics[width=12cm,height=8cm,keepaspectratio]{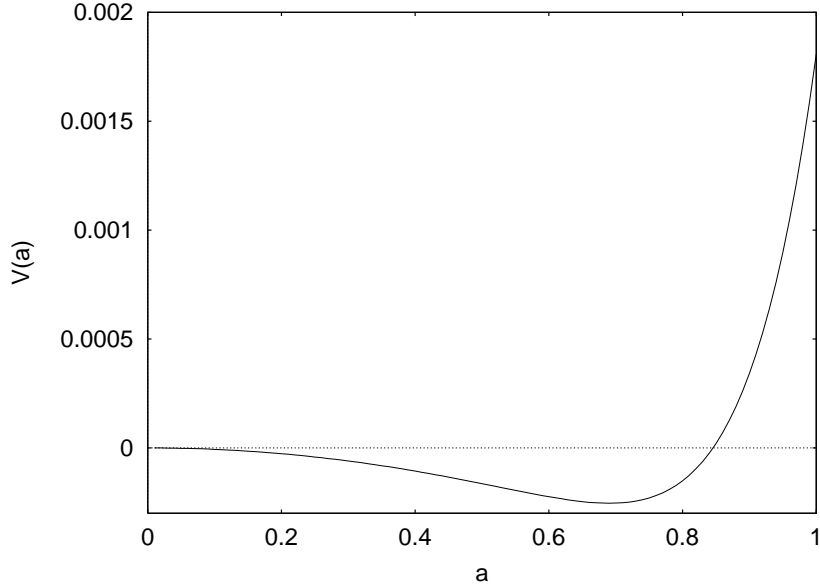}
\end{center}
\caption{The effective potential with the same parameters as in
  Fig.~\ref{EffPot} in the small-$a$ classically allowed region.}
\label{EffPotSmall}
\end{figure}

The presence of a matter potential then leads to a small classically
allowed region between $a=0$ and some positive value (see
Fig.~\ref{EffPotSmall}) because the potential term increases faster
than the suppression in the intrinsic curvature term owing to the
large power in the small-$a$ approximation (\ref{dj}). Initially, the
effective potential decreases implying an inflationary epoch. After
some time, the increasing $\Gamma$ will start to dominate the
effective potential which terminates inflation. When $\Gamma$
increases such that the effective potential reaches positive values,
the classically allowed region ends and a potential hill emerges which
has to be tunneled through by the wave function. After the hill, we
have the second phase of inflation already observed above which
usually will only last for a small number of $e$-foldings. However,
the first phase of inflation can start at values of the scale factor
arbitrarily close to zero such that the amount of inflation can be
very high just as in the flat case.

So far we assumed that the ambiguity parameter $j_{\Gamma}$ in the
intrinsic curvature term is smaller than the parameter $j_{\phi}$ we had
originally in matter terms. If this is not the case, the potential
hill can even disappear completely if $j_{\Gamma}> j_{\phi}$ which implies
a single phase of inflation. The closed model is then very similar to
the flat model for small volume due to the suppression of the
intrinsic curvature term. 

\begin{figure}[ht]
\begin{center}
 \includegraphics[width=12cm,height=8cm,keepaspectratio]{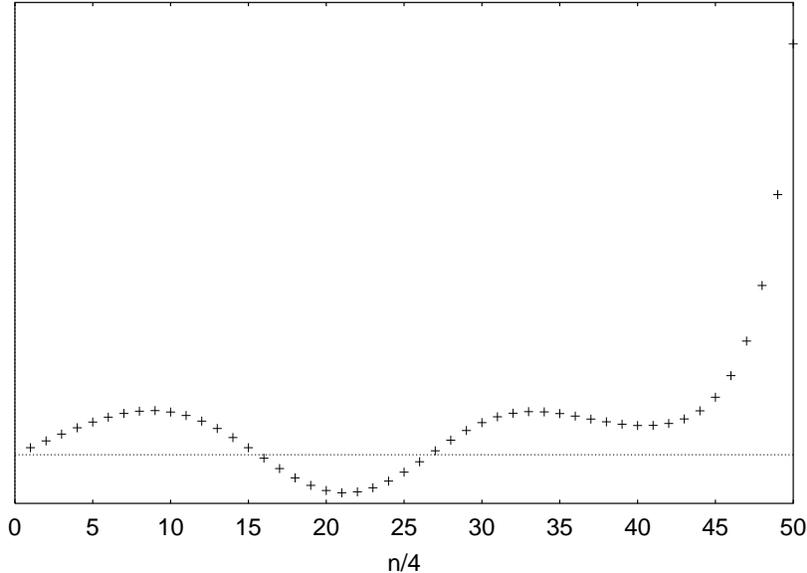}
\end{center}
\caption{Wave function in the small-$a$ classically allowed region
  with $\Gamma$-suppression for $j_{\Gamma}=50$. (Obtained as a
  solution to equation (\ref{Diff}) with
  $4\gamma^2\Gamma_{j_{\Gamma},n}$ replacing $\gamma^2$ on the left
  hand side.)}
\label{WaveGam}
\end{figure}

\begin{figure}[ht]
\begin{center}
 \includegraphics[width=12cm,height=8cm,keepaspectratio]{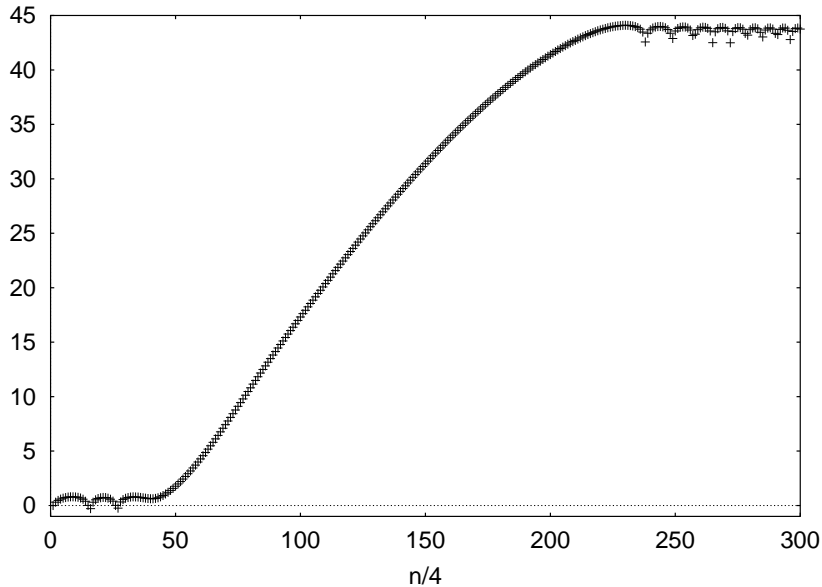}
\end{center}
\caption{Logarithm of the absolute value of the wave function between
  the first two classically allowed regions ($j_{\Gamma}=50$). In the
  classically forbidden region the wave function grows by more than 40
  orders of magnitude and is very similar to a solution without
  $\Gamma$-suppression.}
\label{WaveGamLog}
\end{figure}

Another consequence of the suppression is that the wave function,
which has to start with a small value at small $n$, oscillates in the
first classically allowed region and can grow before it tunnels into
the second region at large volume (Fig.~\ref{WaveGam}). One could
think that this scenario would lead to more similarities between the
discrete wave function and that obtained with the tunneling proposal,
which would be the case if the wave function decays in the classically
forbidden region \cite{tunnelinginfl}. However, in the classically
forbidden region we still have two independent solutions, one
exponentially increasing and one exponentially decreasing. Since the
dynamical initial condition, unlike the tunneling proposal, is not
tailored to select the decreasing part, the increasing part will be
present and dominate the solution (Fig.~\ref{WaveGamLog}).

\subsection{Horizon Problem}

We now seek to answer quantitatively the question of whether or not
enough inflation is obtained with the quantum modifications. A central
problem with the Standard Big Bang (SBB) proposal is the `horizon
problem' in which the comoving region observed by the cosmological
microwave background (CMB) is larger than the comoving forward light
cone at the recombination time $t_{{\rm rec}}$ thus implying regions
of the CMB out of causal contact. The forward light cone $l_{\rm f}
(t_{{\rm rec}})$ is given by
\begin{equation}
 l_{\rm f}(t_{{\rm rec}}) = \int_0^{t_{{\rm rec}}} \frac{1}{a(t)} \md t =
 \int_{a_{{\rm min}}}^{a_{{\rm rec}}} \frac{\md a}{a \dot{a}}
\end{equation}
We consider matter in the form of a scalar field and the Friedmann
equation becomes
\begin{equation}\label{FRW}
\left( \frac{\dot{a}}{a} \right)^2 =\frac{\kappa d_{j_{\phi}}(a)
                                    p_{\phi}^2}{3 a^3}
                                    +\frac{2\kappa}{3} V(\phi)
                                    +\frac{4(\Gamma^2-\Gamma)}{a^2}
\end{equation}
where $p_{\phi}=a^3 \dot{\phi}$ is the conjugate momentum to the
scalar field and $\Gamma$ is the spin connection. We note that when
$\Gamma=1/2$ we regain the usual $-1/a^2$ term of the Friedmann
equation for the closed model. For small $a$ we can use the expansion
for $d_j (a)$ in (\ref{dj}) to derive an approximate expression for
$l_{\rm f}(t_{{\rm rec}})$.

First we consider a free, massless scalar field where the potential is
zero and the conjugate momentum is constant. In the flat model the
curvature term is zero and thus $\dot{a} \sim a^{11/2}$ and the
forward light cone diverges as the integral is taken to small values
of $a$. In the closed model, however, the integral begins at the
minimum classical scale factor such that it does not diverge.  Given
that the amount of inflation is small for the closed model in this
region, it would require very large values for $p_{\phi}$ to overcome
the horizon problem. We will see that this problem is further
evidenced in the flatness problem.

In the presence of a scalar potential, for small enough $a$ the
potential term will dominate over the kinetic term. We have shown that
with $\Gamma$-suppression, the closed model behaves similarly to the
flat model in the small $a$ limit. Since the kinetic term is
suppressed, the scalar field remains nearly constant and standard
inflationary behavior occurs close to the classical
singularity. There, $a$ grows exponentially and $\dot{a} \sim a$, thus
the forward light cone diverges beginning at the classical
singularity.

\subsection{Flatness Problem}

The flatness problem represents itself as a fine-tuning problem. The
dimensionless parameter $\Omega = \rho/\rho_{\rm c} =
\kappa\rho/3H^2$, where $\rho_{\rm c}= 3 H^2/\kappa$ is the critical
density and $H=\dot{a}/a$ is the Hubble parameter, indicates the
spatial topology: $\Omega>1$ gives a closed universe whereas
$\Omega\le1$ gives an open universe. In the SBB model, $\Omega=1$ is
an unstable fixed point; the deviation $\epsilon=\Omega-1$ grows like
$a$ for a matter dominated universe and $a^2$ for a radiation
dominated universe. Current measurements put a value of $\Omega
\approx1$ \cite{Omega} requiring $\Omega$ to be extraordinarily close
to one in the early universe in the SBB scenario.

We consider $\epsilon=\Omega-1=k/\dot{a}^2$ and use the Friedmann 
equation (\ref{FRW}) to write $\epsilon$ as a function of the scale
factor $a$
\begin{equation}\label{epsilon}
\epsilon=k\left(\case{1}{3}\kappa d_{j_{\phi}}(a)a^{-1} p_{\phi}^2
                +\case{2}{3}\kappa a^2V(\phi) -k\right)^{-1}.
\end{equation}
In standard inflation, the universe undergoes a period of exponential
growth in which $\epsilon$ is driven close to zero (spatial flatness)
thus avoiding the necessity of fine-tuning $\Omega$ and predicting a
current value of $\Omega$ close to one for most models. With quantum
modifications $\epsilon$ is also driven toward zero during the
inflationary period. Considering a massless scalar field, for small
$a$ the kinetic term is an increasing function of $a$ and thus
$\epsilon$ decreases until $d_{j_{\phi}}(a)$ becomes a decreasing
function after which $\epsilon$ grows with $a$. The minimum value of
$\epsilon$ can be approximated by using the fact that
$d_{j_{\phi}}(a)$ takes its maximum value at $a_{\rm
peak}\approx\sqrt{\gamma \lP^2 j_{\phi}/3}$ with $d_{j, {\rm max}}
\approx a_{\rm peak}^{-3}$ \cite{Ambig}. This gives a minimum value
\begin{equation}
\epsilon_{{\rm min}} \approx \left( \frac{\kappa p_{\phi}^2}{3} 
		\left(\frac{3}{\gamma \lP^2j_{\phi}}\right)^2-1 \right)^{-1}.
\end{equation}
Assuming there exists a region where the matter kinetic term dominates
the curvature term we find that $\epsilon_{{\rm min}} \sim
j_{\phi}^2/p_{\phi}^2$, thus in order to drive $\epsilon$ sufficiently
close to zero we need small values of $j_{\phi}$ and large values of
$p_{\phi}$. This presents a problem since $j_{\phi}$ needs to be
sufficiently large so that the quantum corrections occur in the region
where the classical equations of motion are valid.  If we take a
reasonable value of $j_{\phi}=100$ we find that in order for
$\epsilon$ to have the required accuracy of $10^{-60}$ near the Plank
regime (assuming no second round of standard inflation),
$\sqrt{\kappa}p_{\phi}$ needs to be on the order of $10^{30}\lP^2$. It
thus seems unlikely that the quantum corrections alone can provide
enough inflation for the closed model.

The inclusion of a potential does not help the matter. To drive
$\epsilon$ close enough to zero would require either unnaturally large
values of the potential or very large values of the final scale factor
at the end of inflation. The latter scenario would then be equivalent
to standard inflation.

\subsection{Summary}

We have seen that the ratio $a_{\rm f}/a_{\rm i}$ cannot be large
enough for a sufficient amount of inflation with natural values of the
different parameters, which are mainly the half-integer ambiguity
label $j_{\phi}$ and the initial value $p_{\phi}$ of the scalar
momentum. As a function of these parameters, $a_{\rm f}/a_{\rm i}$
increases only slowly with $p_{\phi}$ and decreases with $j_{\phi}$
(small values of $j_{\phi}$ also appear more natural from a conceptual
point of view). If we invoke $\Gamma$-suppression introducing a new
parameter $j_{\Gamma}$, the situation looks better since we have an
early inflationary region with small $a_{\rm i}$. However, in this regime
the viability of the effective framework remains to be understood.

\begin{figure}[ht]
\begin{center}
 \includegraphics[width=12cm,height=6cm,keepaspectratio]{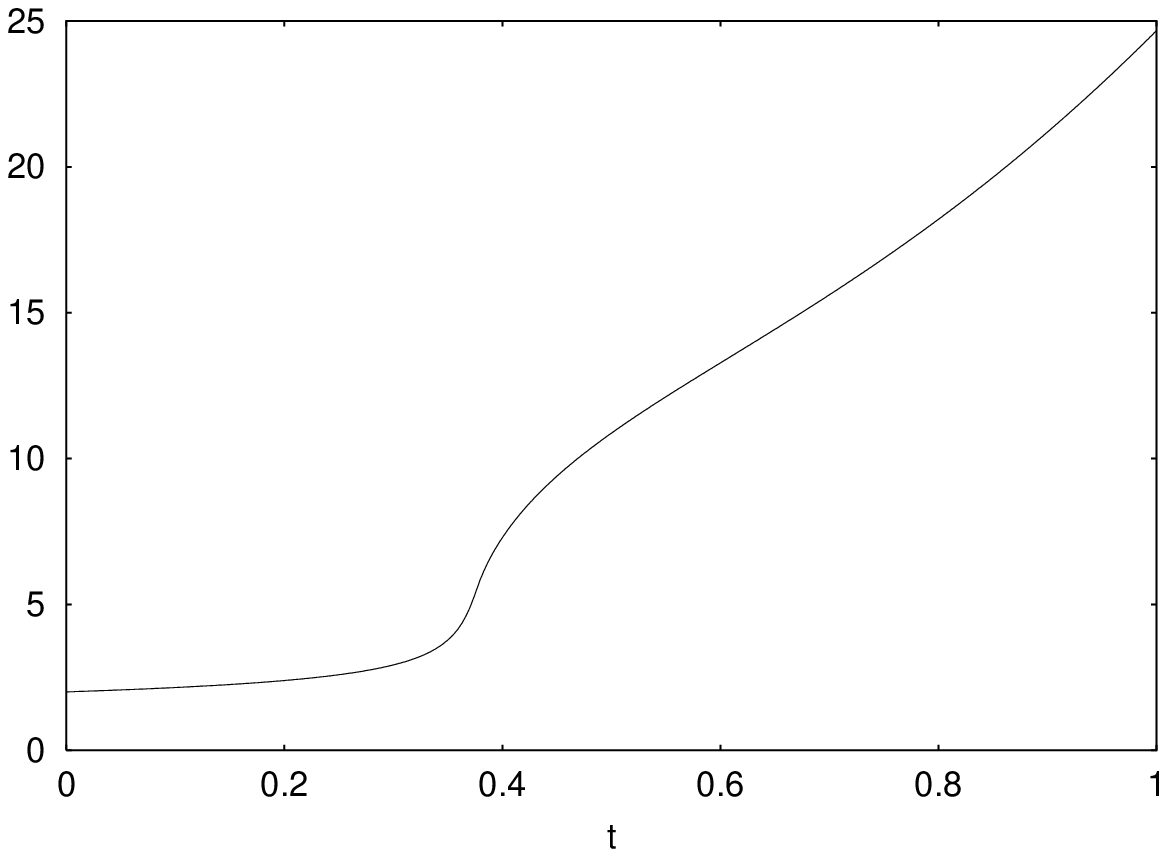}
 \includegraphics[width=12cm,height=6cm,keepaspectratio]{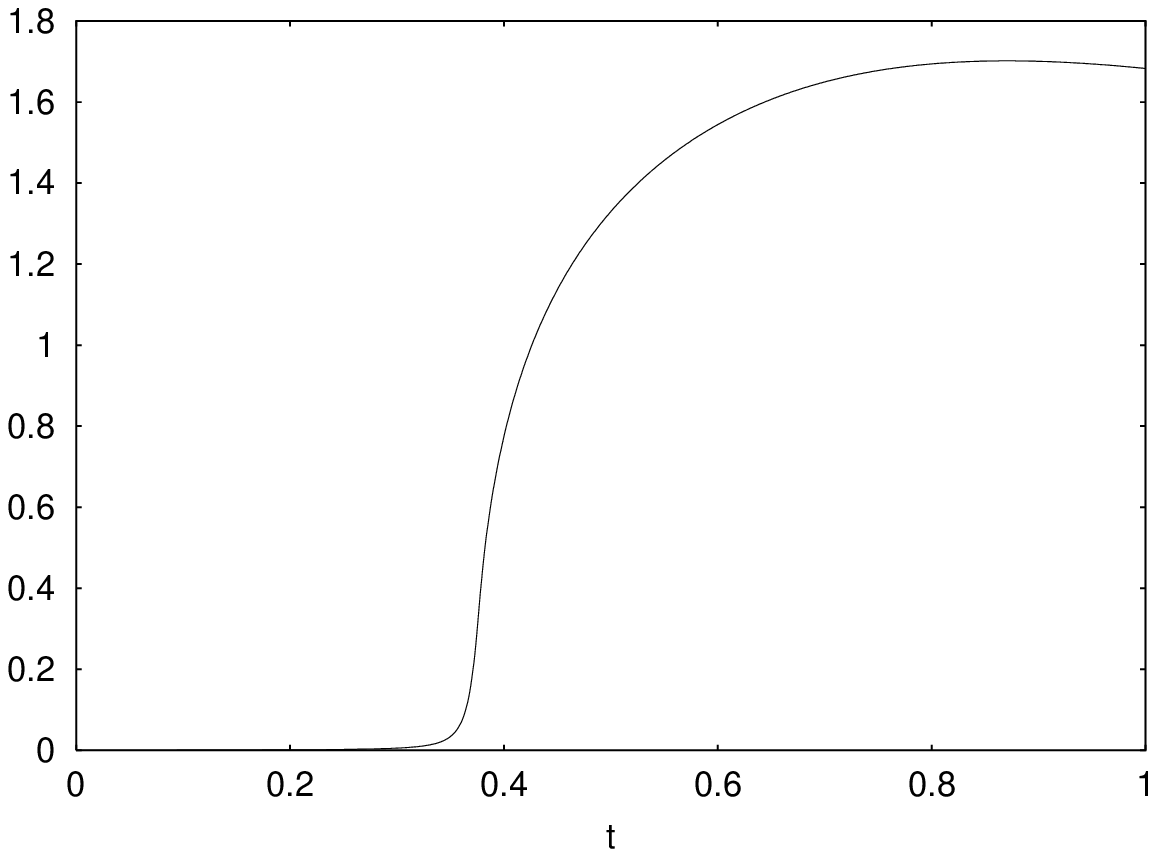}
\end{center}
\caption{Scale factor $a$ (top) and scalar $\phi$ (bottom) in Planck
  units with a mass term $\kappa V(\phi)=10^{-3}\hbar\phi^2/2$. The
  first phase of quantum geometry inflation (which ends at $t\approx
  0.4$) leads to large initial $\phi$ for a second phase of slow roll
  inflation (with $j_{\phi}=100$ and initial values $\phi_0=0$,
  $\sqrt{\kappa}\dot{\phi}_0\approx 10^{-5}\lP^{-1}$ at $a_0=2\lP$
  such that $\sqrt{\kappa}p_{\phi,0}\approx 100\lP^2$).}
\label{SlowRoll}
\end{figure}

The best scenario can be obtained by using quantum geometry inflation
in order to generate standard inflation. This happens naturally
whenever there is a matter component which is well approximated by a
scalar with a flat potential. In the quantum geometry regime the
equations of motion of the scalar have a different form than usually
because of the modified density. Equations of motion can be derived
from the Hamiltonian which for a scalar has the effective form
\[
 H_{\rm matter}^{\rm eff}(a)= \frac{1}{2}
 d_{j_{\phi}}(a)p_{\phi}^2+a^3 V(\phi)\,.
\]
This yields
\[
 \dot{\phi}=\{\phi,H_{\rm matter}^{\rm eff}(a)\}= d_{j_{\phi}}(a)p_{\phi}
\]
and
\[
 \dot{p}_{\phi}=\{p_{\phi},H_{\rm matter}^{\rm eff}(a)\}= -a^3 V'(\phi)
\]
resulting in the second order equation of motion
\[
 \ddot{\phi}=p_{\phi}\frac{\md d_{j_{\phi}}(a)}{\md
 t}+d_{j_{\phi}}(a)\dot{p}_{\phi}=
 p_{\phi}\,\dot{a}\,d_{j_{\phi}}'(a)- a^3 d_{j_{\phi}}(a)V'(\phi)=
 a\frac{\md\log d_{j_{\phi}}(a)}{\md a}\, H\dot{\phi}- a^3
 d_{j_{\phi}}(a)V'(\phi)
\]
with the Hubble parameter $H$.  In the standard case $d_j(a)=a^{-3}$,
we have $a\,\md\log d_j(a)/\md a=-3$ and the first order term serves
as a friction which leads to slow roll for a sufficiently flat
potential. For the modified $d_j(a)$ at small volume, however,
$d_j(a)$ increases and thus $a\,\md\log d_j(a)/\md a$ is positive. In
this case, the friction term has the opposite sign and forces the
scalar to move up the potential (see Fig.~\ref{SlowRoll}, whose
initial values have been chosen for the purpose of illustration; one
can easily achieve larger maximal values of $\phi$ by using a larger
initial $\dot{\phi}$).

Thus, quantum geometry modifications drive the scalar up the
potential during early phases, even if it starts in a minimum, thereby
producing large initial values for standard inflation. It remains to
be seen whether the standard inflationary phase will wash away any
possible signature of the quantum geometry phase, or if it can be
distinguished from other scenarios.

\section{Discussion}

We have studied the closed isotropic model in loop quantum cosmology
and seen that it reproduces all of the main general results: its
evolution is non-singular and it predicts dynamical initial conditions
as well as the occurrence of inflation. The advantage of using the
closed model is that we are able to compare with older attempts to
understand the quantum dynamics. In fact, those approaches are
reproduced at large volume, but many new properties result at small
volume. The initial conditions, for instance, have a very different
origin since they are derived in part from the dynamical law and not
imposed by hand. Furthermore, the factor ordering of the constraint is
fixed by the requirement of a non-singular evolution. Older attempts
have often been plagued by the factor ordering problem because
different orderings can lead to qualitatively different results and
even invalidate certain proposals \cite{Konto}. Still, as in any
quantization there are many ambiguities which can be included with
certain parameters like $j_{\phi}$. These ambiguities, however,
usually lead only to quantitative differences leaving the main results
unchanged. They can, therefore, be used for a phenomenological
analysis.

As for dynamical initial conditions, the constraint equation requires
the wave function to approach very small values at the classical
singularity such that the effect is often similar to DeWitt's initial
condition. However, due to effects of the discreteness the initial
value problem is well-posed (which presents some realization of the
Planck potential of \cite{SIC}). The first quantization we used for
the closed model is closely related to the no-boundary proposal while
it would be different from the tunneling proposal (we note, however,
that we ignored the kinetic term of the matter Hamiltonian in
(\ref{deSitter}) as usual in this context; in a more general
discussion one can also expect more differences to the no-boundary
proposal due to quantum geometry modifications of the kinetic term as
in \cite{Inflation}). However, this quantization does not take into
account a possible suppression of intrinsic curvature which can be
realized by embedding the isotropic model in the anisotropic Bianchi
IX model. This introduces a second ambiguity parameter $j_{\Gamma}$
which influences the results: If $j_{\phi} < j_{\Gamma}$, there is no
potential hill and the whole region of scale factors between zero and
a possible recollapse value is allowed classically. In this case, the
closed model is very similar to the flat model at small
volume. Otherwise, there will be a potential hill and we have two
classically allowed regions. For small $j_{\Gamma}$, the first region
is only small and the wave function again resembles that of the
no-boundary proposal. If $j_{\Gamma}$ is increased, the first
classically allowed region grows such that the wave function can
increase there before it tunnels into the second classically allowed
region. However, the result remains closer to the no-boundary proposal
than to the tunneling proposal because the exponentially increasing
branch of the wave function will dominate in the classically forbidden
region. Still, depending on the values of ambiguity parameters we
obtain modifications of the old proposals realizing different aspects
of them. Furthermore, the values will influence the amount of
inflation achieved within a given model.

We also have to give some cautionary remarks: While the quantization
of the flat model can be regarded as being very close to that of the
full theory, the closed model requires a special input due to the
large intrinsic curvature. In the flat model, the intrinsic curvature
vanishes identically, and in the full theory it can be ignored
locally. Therefore, it has to be verified that the physical effects
are not artifacts of the special techniques but indeed model the full
theory. As we have seen here, viewing a symmetric model as being
embedded in a less symmetric one can lead to important effects. In our
case, those effects ($\Gamma$-suppression) were welcome and improved
the model from the point of view of cosmological model building. It is
certainly a very important issue to investigate a similar, but much
more complicated, correspondence between symmetric models and the full
theory.  Also more complicated models contain many ambiguity
parameters whose physical role and reasonable values are not
completely clear yet. Still, they provide a rich ground for
phenomenology which allows studying the same objects and effects
present in the full theory in a much more simplified context.

\section*{Acknowledgments}

We are grateful to A.~Ashtekar for discussions and to A.~Vilenkin for
discussions and for hospitality to one of us (M.B.). This work was
supported in part by NSF grant PHY00-90091 and the Eberly research
funds of Penn State.


\begin{thebibliography}{10}

\bibitem{DeWitt}
B.~S.\ DeWitt,
\newblock Quantum Theory of Gravity. I. The Canonical Theory,
\newblock {\em Phys.\ Rev.} 160 (1967) 1113--1148

\bibitem{Misner}
C.~W.\ Misner,
\newblock Quantum Cosmology. I,
\newblock {\em Phys.\ Rev.} 186 (1969) 1319--1327

\bibitem{nobound}
J.~B.\ Hartle and S.~W.\ Hawking,
\newblock Wave function of the Universe,
\newblock {\em Phys.\ Rev.\ D} 28 (1983) 2960--2975

\bibitem{tunneling}
A.\ Vilenkin,
\newblock Quantum creation of universes,
\newblock {\em Phys.\ Rev.\ D} 30 (1984) 509--511

\bibitem{Nonpert}
A.\ Ashtekar,
\newblock {\em Lectures on non-perturbative canonical gravity},
\newblock World Scientific, Singapore, 1991

\bibitem{Rov:Loops}
C.\ Rovelli,
\newblock Loop Quantum Gravity,
\newblock {\em Living Reviews in Relativity} 1 (1998)
  http://www.livingreviews.org/Articles/Volume1/1998--1rovelli, [gr-qc/9710008]

\bibitem{ThomasRev}
T.\ Thiemann,
\newblock Introduction to Modern Canonical Quantum General Relativity,
\newblock gr-qc/0110034

\bibitem{SymmRed}
M.\ Bojowald and H.~A.\ Kastrup,
\newblock Symmetry Reduction for Quantized Diffeomorphism Invariant Theories of
  Connections,
\newblock {\em Class.\ Quantum Grav.} 17 (2000) 3009--3043, [hep-th/9907042]

\bibitem{cosmoI}
M.\ Bojowald,
\newblock Loop Quantum Cosmology: I. Kinematics,
\newblock {\em Class.\ Quantum Grav.} 17 (2000) 1489--1508, [gr-qc/9910103]

\bibitem{IsoCosmo}
M.\ Bojowald,
\newblock Isotropic Loop Quantum Cosmology,
\newblock {\em Class.\ Quantum Grav.} 19 (2002) 2717--2741, [gr-qc/0202077]

\bibitem{Sing}
M.\ Bojowald,
\newblock Absence of a Singularity in Loop Quantum Cosmology,
\newblock {\em Phys.\ Rev.\ Lett.} 86 (2001) 5227--5230, [gr-qc/0102069]

\bibitem{DynIn}
M.\ Bojowald,
\newblock Dynamical Initial Conditions in Quantum Cosmology,
\newblock {\em Phys.\ Rev.\ Lett.} 87 (2001) 121301, [gr-qc/0104072]

\bibitem{Inflation}
M.\ Bojowald,
\newblock Inflation from quantum geometry,
\newblock {\em Phys.\ Rev.\ Lett.} 89 (2002) 261301, [gr-qc/0206054]

\bibitem{Scalar}
M.\ Bojowald and F.\ Hinterleitner,
\newblock Isotropic loop quantum cosmology with matter,
\newblock {\em Phys.\ Rev.\ D} 66 (2002) 104003, [gr-qc/0207038]

\bibitem{Spin}
M.\ Bojowald, G.\ Date, and K.\ Vandersloot,
\newblock Homogeneous loop quantum cosmology: The role of the spin connection,
\newblock in preparation

\bibitem{AshVar}
A.\ Ashtekar,
\newblock New Hamiltonian Formulation of General Relativity,
\newblock {\em Phys.\ Rev.\ D} 36 (1987) 1587--1602

\bibitem{AshVarReell}
J.~F.\ Barbero~G.,
\newblock Real Ashtekar Variables for Lorentzian Signature Space-Times,
\newblock {\em Phys.\ Rev.\ D} 51 (1995) 5507--5510, [gr-qc/9410014]

\bibitem{Bohr}
A.\ Ashtekar and M.\ Bojowald,
\newblock Mathematical structure of loop quantum cosmology,
\newblock in preparation

\bibitem{Immirzi}
G.\ Immirzi,
\newblock Real and Complex Connections for Canonical Gravity,
\newblock {\em Class.\ Quantum Grav.} 14 (1997) L177--L181

\bibitem{QSDV}
T.\ Thiemann,
\newblock {QSD V}: Quantum Gravity as the Natural Regulator of Matter Quantum
  Field Theories,
\newblock {\em Class.\ Quantum Grav.} 15 (1998) 1281--1314, [gr-qc/9705019]

\bibitem{InvScale}
M.\ Bojowald,
\newblock Inverse Scale Factor in Isotropic Quantum Geometry,
\newblock {\em Phys.\ Rev.\ D} 64 (2001) 084018, [gr-qc/0105067]

\bibitem{Ambig}
M.\ Bojowald,
\newblock Quantization ambiguities in isotropic quantum geometry,
\newblock {\em Class.\ Quantum Grav.} 19 (2002) 5113--5130, [gr-qc/0206053]

\bibitem{QSDI}
T.\ Thiemann,
\newblock Quantum Spin Dynamics {(QSD)},
\newblock {\em Class.\ Quantum Grav.} 15 (1998) 839--873, [gr-qc/9606089]

\bibitem{cosmoIII}
M.\ Bojowald,
\newblock Loop Quantum Cosmology III: Wheeler-DeWitt Operators,
\newblock {\em Class.\ Quantum Grav.} 18 (2001) 1055--1070, [gr-qc/0008052]

\bibitem{SemiClass}
M.\ Bojowald,
\newblock The Semiclassical Limit of Loop Quantum Cosmology,
\newblock {\em Class.\ Quantum Grav.} 18 (2001) L109--L116, [gr-qc/0105113]

\bibitem{HomCosmo}
M.\ Bojowald,
\newblock Homogeneous loop quantum cosmology,
\newblock gr-qc/0303073

\bibitem{Konto}
N.\ Kontoleon and D.~L.\ Wiltshire,
\newblock Operator ordering and consistency of the wavefunction of the
  Universe,
\newblock {\em Phys.\ Rev.\ D} 59 (1999) 063513, [gr-qc/9807075]

\bibitem{tunnelinginfl}
A.\ Vilenkin,
\newblock Quantum cosmology and the initial state of the Universe,
\newblock {\em Phys.\ Rev.\ D} 37 (1988) 888--897

\bibitem{Omega}
C.~L.\ Bennett and et~al.,
\newblock First Year Wilkinson Microwave Anisotropy Probe (WMAP) Observations:
  Preliminary Maps and Basic Results, [astro-ph/0302207],
\newblock see also
  \verb|http://map.gsfc.nasa.gov/m_mm/pub_papers/firstyear.html|

\bibitem{SIC}
H.~D.\ Conradi and H.~D.\ Zeh,
\newblock Quantum cosmology as an initial value problem,
\newblock {\em Phys.\ Lett.\ A} 154 (1991) 321--326

\end{thebibliography}

\end{document}